\documentclass[
]{ceurart}

\usepackage{listings}
\usepackage{multirow}
\usepackage{arydshln}
\usepackage{graphicx}  
\begin{document}

\copyrightyear{2025}
\copyrightclause{Copyright for this paper by its authors. Use permitted under Creative Commons License Attribution 4.0 International (CC BY 4.0)}

\conference{SIGIR 2025: Workshop on eCommerce, June 17, 2025, Padua, Italy}


\title{Intent-Aware Neural Query Reformulation for Behavior-Aligned Product Search}




\author[1]{Jayanth Yetukuri}[email=jyetukuri@ebay.com]
\author[1]{Ishita Khan}[email=ishikhan@ebay.com]
\address[1]{eBay Inc, San Jose, CA, USA}


\begin{abstract}
Understanding and modeling buyer intent is a foundational challenge in optimizing search query reformulation within the dynamic landscape of e-commerce search systems. This work introduces a robust data pipeline designed to mine and analyze large-scale buyer query logs, with a focus on extracting fine-grained intent signals from both explicit interactions and implicit behavioral cues. Leveraging advanced sequence mining techniques and supervised learning models, the pipeline systematically captures patterns indicative of latent purchase intent, enabling the construction of a high-fidelity, intent-rich dataset. 

The proposed framework facilitates the development of adaptive query rewrite strategies by grounding reformulations in inferred user intent rather than surface-level lexical signals. This alignment between query rewriting and underlying user objectives enhances both retrieval relevance and downstream engagement metrics. Empirical evaluations across multiple product verticals demonstrate measurable gains in precision-oriented relevance metrics, underscoring the efficacy of intent-aware reformulation. Our findings highlight the value of intent-centric modeling in bridging the gap between sparse user inputs and complex product discovery goals, and establish a scalable foundation for future research in user-aligned neural retrieval and ranking systems.
\end{abstract}


\maketitle

\section{Introduction}
\label{SEC:INTRO}
The rapid growth of e-commerce has been fueled by advancements in technology and increasing consumer trust in online platforms. Innovations in logistics, mobile accessibility, digital payments, and personalized marketing have all contributed to making online shopping more seamless and appealing. Among the many touch-points in a buyer's journey, the single most frequent and critical point of interaction for a buyer is typically a \textit{search bar}. This interface serves as the primary conduit through which buyers communicate their needs and expectations. Here, buyers specify their intent through a keyword query, which consists of a sequence of words, or tokens, designed to express their purchase intent. These queries may range from generic (e.g., ``laptop'') to a highly specific (e.g., ``13-inch MacBook Pro M2 2023''), depending on the user's familiarity with the product domain and clarity of need.

This query acts as the trigger for the search engine \cite{Trotman2017TheAO}, which employs a series of sophisticated processes, increasingly powered by Artificial Intelligence (AI) models, to generate a Search Results Page (SRP). These processes include query understanding, retrieval, ranking, and often re-ranking based on user behavior signals and contextual factors. The SRP includes a relevant recall: a curated list of items retrieved from the inventory database, tailored to match the user's intent and guide them towards a potential purchase. A well-optimized SRP is critical not only for helping users find what they are looking for efficiently but also for driving key business metrics such as click-through rate, and  conversion rate.

\begin{figure}
    \includegraphics[width=1\linewidth, trim=0 4cm 0.5cm 1cm, clip]{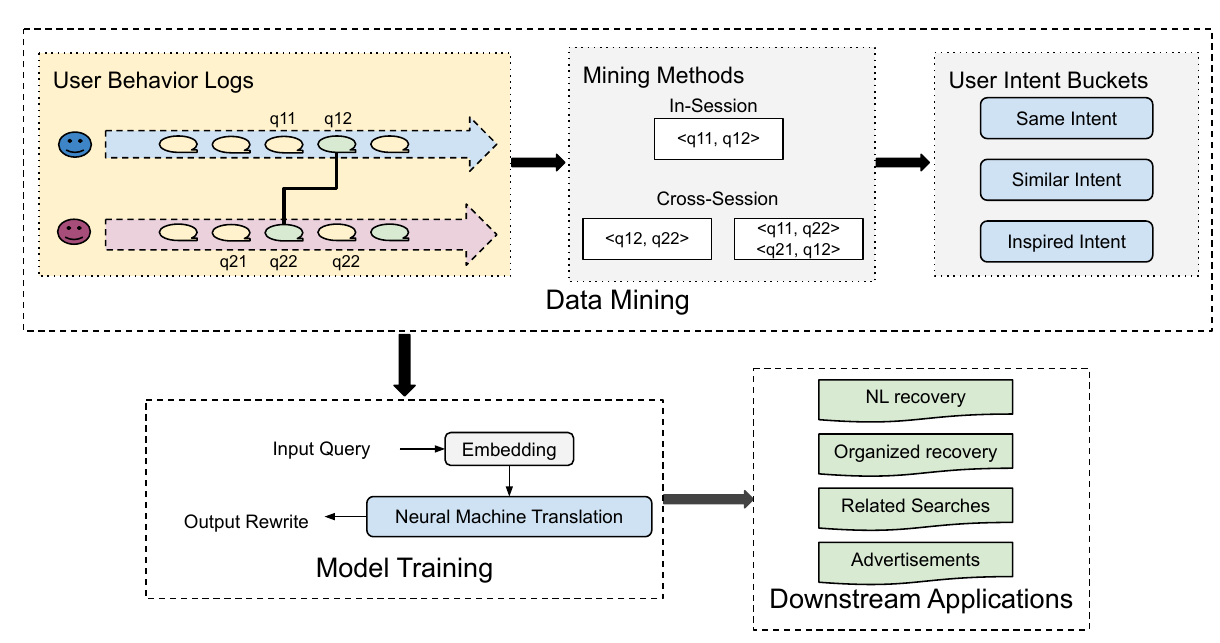}
    \caption{Overview of Search Engine components and different modules which leverage query reformulation capability.}
    \label{FIG:OVERVIEW}
\end{figure}

A central challenge in the domain of neural search and retrieval, particularly within e-commerce environments, is the accurate semantic interpretation of user-issued queries, which are often short, ambiguous, or lexically sparse. As demonstrated in prior work \cite{yetukuri2024, ecommsurvey}, models that anticipate user intent by predicting downstream behaviors—such as product engagement or purchase—enable the generation of semantically aligned \textit{alternative query formulations}. These reformulations, grounded in behavioral modeling and intent inference, have shown to significantly enhance retrieval relevance and user satisfaction \cite{riezler2010query}. Such advancements underscore the growing role of behavior-informed query rewriting frameworks as core components in next-generation search architectures.

Recent efforts, including \cite{Tan2017QueryRF}, have addressed a particularly pressing issue: queries yielding limited or null recall due to lexical overspecification, inventory sparsity, or long-tail item structures. Traditional keyword-based retrieval systems are often ill-equipped to handle such queries. By integrating sequence-to-sequence (seq2seq) architectures, researchers have developed methods to rephrase queries into semantically equivalent but inventory-aligned alternatives. This not only mitigates search failure rates but also improves system robustness in cold-start and long-tail scenarios. Crucially, this line of work highlights the need for reformulation systems that are both context-aware and inventory-sensitive, effectively bridging the semantic gap between user intent and system recall \cite{Analyzing,Patterns,Intent,Structured,awadallah2013beyond}.

Building on this foundation, Yetukuri et al. \cite{yetukuri2024} extended the paradigm further by introducing a counterfactual reasoning framework for query reformulation. This approach conceptualizes reformulation as a means of answering "what-if" scenarios—proposing plausible, intent-preserving alternatives to queries that underperform due to structural or semantic misalignments with the available catalog. By learning from historical user behavior and failed retrieval instances, the model constructs counterfactual versions of queries that simulate higher-engagement trajectories. This allows for preemptive corrections in cases of null or low-recall queries \cite{mandal2019query,Trotman2017TheAO,huangRelevant}, offering not just higher recall but a principled method for evaluating the efficacy of reformulated queries across diverse catalog and intent distributions.

Collectively, these approaches represent a significant shift toward intent-centric and behavior-informed retrieval systems. They demonstrate the viability of deep learning models that do more than match tokens—they anticipate goals, model affordances, and optimize outcomes through semantic and behavioral inference, laying a path forward for intelligent, adaptive, and user-aligned search engines in e-commerce.

\subsection{Intent-Aware Query Reformulation Framework}

To advance the state of the art in query reformulation for e-commerce search, our ongoing research centers on the development of a \textit{comprehensive, intent-aware modeling framework} that tailors query rewriting strategies to distinct buyer intent categories. Specifically, we operationalize buyer queries into canonical intent classes—\textit{transactional}, \textit{navigational}, and \textit{informational}—and apply differentiated reformulation mechanisms optimized for each intent bucket.

A central component of this framework is the construction of high-quality, labeled training data that reflects the nuanced behavioral semantics of user intent. We introduce three intent-aligned query transformation types: \textit{Same Intent}, \textit{Similar Intent}, and \textit{Inspired Intent}, which capture varying degrees of semantic proximity and behavioral continuity. These labels are derived through a hybrid approach, leveraging both \textit{in-session} and \textit{cross-session} query trajectories mined from real-world search logs. This enables the model to learn from both direct reformulation patterns and longer-term buyer journeys. The proposed system is designed to support multiple downstream applications within e-commerce search pipelines, including:

\begin{enumerate}
    \item \textbf{Recall Enhancement in Low Inventory Scenarios:}
    Intent-specific reformulations help mitigate low-recall issues by broadening lexical coverage and semantically aligning user queries with available inventory.
    \item \textbf{Query Recommendation for Related Search Exploration:}
    Reformulated queries act as guided pathways toward related or complementary searches, promoting deeper user engagement and increasing the likelihood of conversion.
    \item \textbf{Item Recommendations at the Bottom of the Search Results Page (BOS):}
    Inspired intent reformulations facilitate the generation of diverse, intent-divergent recommendations that extend the scope of the original query, enriching the tail-end of the SRP with broader discovery opportunities.
\end{enumerate}

By unifying these components into an intent-conditioned reformulation framework, our approach aims to deliver not only precision-enhanced retrieval but also a contextually adaptive and behaviorally informed search experience. This research opens new avenues in personalization, advanced query understanding, and session-aware ranking strategies within e-commerce systems.

This comprehensive framework leverages the synergy between intent inference and reformulation generation to address multiple pain points in modern search pipelines. By integrating these strategies into production-grade e-commerce platforms, we aim to enable a robust, AI-driven search stack that enhances retrieval coverage, improves relevance, and drives measurable gains in customer satisfaction and business KPIs. The novelty of this framework lies in three core dimensions:

\begin{table}
    \centering
    \caption{Real-world examples collected into three buckets.}
    \label{TAB:DATASET-EXAMPELES}
    \begin{tabular}{lcc}
    \toprule
        \textbf{Bucket}&\textbf{Source query}&{\textbf{Target query}}\\
    \midrule
        \multirow{3}{1.4cm}{\emph{Same Intent}}&{nike air jordan 4}&{nike air jordan 11}\\
        &{iphone 11 unlocked}&{iphone 14 unlocked}\\
        &{maga hat official}&{maga hat authentic}\\
        
        \cmidrule(lr){2-3}
        \multirow{3}{1.4cm}{\emph{Similar Intent}}&{nike womens size 9}&{nike womens air max size 9}\\
        &{iphone 14 plus case}&{phone case with stand}\\
        &{kobe bryant nike authentic jersey}&{michael jordan authentic jersey 44}\\
        
        \cmidrule(lr){2-3}
        \multirow{3}{1.4cm}{\emph{Inspired Intent}}&{sansung filp}&{iphone 14}\\
        &{adidas adios pro 4}&{nike ultrafly trail 12}\\
        &{transformers logo stickers}&{white autobot sticker}\\
    \bottomrule
    \end{tabular}
\end{table}

\begin{enumerate}
    \item \textbf{Intent-Specific Training:}
    The proposed framework facilitates intent-conditioned reformulation strategies explicitly tailored to distinct stages of the buyer journey. This fine-grained modeling allows the system to selectively optimize for specific outcomes—such as improving recall in underperforming query segments (e.g., null or low-recall cases), while simultaneously supporting exploratory behavior through \textit{inspired} reformulations. Such versatility ensures broad applicability across heterogeneous search intents and product verticals.
    \item \textbf{Behaviorally-Driven Reformulation Signals:}
    The model leverages a rich spectrum of behavior-derived signals, including transactional logs, co-clicked pairs, and multi-hop navigation paths, to inform reformulation generation. By grounding data construction in user interaction patterns—rather than relying on rule-based or synthetic alternatives—the system produces semantically robust reformulations that align with actual buyer intent, leading to more relevant and effective search outcomes.
    \item \textbf{Application Versatility Across Search Tasks:}
    The framework is designed for generalization across a diverse array of retrieval and recommendation tasks. It supports both \textit{intent preservation}—ensuring continuity for goal-driven queries—and \textit{intent diversification}—enabling discovery through related, exploratory reformulations. This dual capability allows the system to act as a bridge between traditional retrieval and modern recommendation paradigms, positioning it as a unified solution for recall enhancement and intent-aligned buyer engagement.
\end{enumerate}

\section{Architecture}

The architecture of our system begins with the extraction of user-generated query reformulations from large-scale behavioral data logs. The primary focus is on identifying \textit{successful} reformulations, characterized by transitions from an initial query to a reformulated query that leads to meaningful user engagement on the Search Results Page (SRP), such as clicks or conversions.

To model the diversity of buyer intent, the query mining process is stratified based on distinct behavioral signals. Reformulations that occur within the same user session are interpreted as expressions of the same underlying intent, especially when they converge on similar engagement patterns. To capture semantically aligned but temporally separated reformulations, we mine cross-session query pairs that have shared clicked items, inferring a shared or similar intent. Finally, to address more divergent, exploratory user behavior, we construct inspired intent query pairs by identifying one-hop SRP neighborhood transitions originating from the cross-session, co-clicked query pairs. These different pathways enable the model to capture a spectrum of buyer intent ranging from strict preservation to conceptual diversification. A subsequent post-filtering stage refines the mined query pairs by applying constraints on categorical alignment, buffered recall similarity, and query length compatibility. This step ensures that each pair faithfully represents its intended intent bucket and that the dataset adheres to the conceptual rigor necessary for effective model training. Representative examples of these mined query pairs are illustrated in Table~\ref{TAB:DATASET-EXAMPELES}, demonstrating the qualitative distinctions across intent types.

Model training is performed using a sequence-to-sequence neural machine translation (NMT) model~\cite{neubig2017neural}, situated within a generative language modeling framework. The model is explicitly conditioned on intent through the use of intent-type tags appended to each training instance, enabling it to learn intent-specific reformulation strategies in a multitask learning setup. The resulting model is capable of generating reformulated queries aligned with specific buyer intent categories.

At inference time, the trained model is deployed across multiple search-facing applications. It generates reformulations for queries that exhibit low or null recall, offers related search suggestions to support user discovery, enhances merchandising and sponsored listing coverage, and enables structured recovery on the SRP through horizontal carousels representing different facets of buyer intent. Figure~\ref{FIG:OVERVIEW} illustrates the full-stack system integration, where incoming buyer queries are processed by the reformulation model to produce candidate suggestions across intent categories. Each application downstream applies its own post-processing—such as filtering, re-ranking, or enforcing categorical constraints—to tailor the reformulations to its operational requirements. This architecture supports both precision-oriented and exploratory search behaviors, providing a comprehensive, intent-aware solution for e-commerce retrieval systems.

\section{Methodology}
Buyer driven reformulations provides an immense information about query intent. This is further vouched by engagement signals often captured by the number of items clicked within an SRP corresponding to the query. A User Engagement score captures several signals like \textit{click}, \textit{bought}, \textit{bid} etc. For the purpose of this study, engagement score is considered to be a weighted combination of the available engagement signals. The dataset is collected from $4$ weeks of buyer search logs. In this section we consolidate the techniques used to mine buyer reformulations and the post processing steps required to segregate the query pairs into intent buckets. 
\subsection{Mining Buyer Query Reformulation}
This section consolidates the three unique modules for mining buyer reformulations. 

\begin{figure}
    \centering
    \includegraphics[width=0.65\linewidth, trim=4cm 12cm 6cm 3cm, clip]{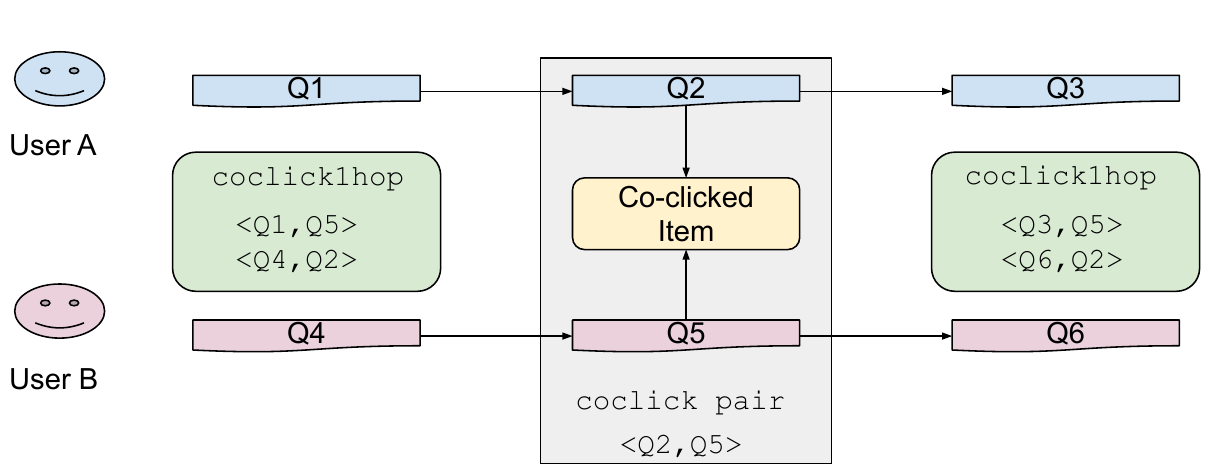}
    \caption{Architecture of mining cross-session buyer rewrites.}    \label{FIG:CROSS-SESSION}
\end{figure}

\paragraph{In-session n-hop buyer query reformulations:} 
Here the focus is on maintaining buyer’s original intent while optimizing recall or aligning it with available inventory. Training data is derived from a source-target query pair, i.e., queries within same session where the source query (initial query) transitions to a target query (reformulated query), and the target query leads to a successful transaction. This decision is reflected in the form of engagement which can be captured as a variety of signals like items clicked, items bought, items added to cart, bidding offers made etc. Buyer reformulation data mining is primarily driven by this method, which causes \textit{Contextual bias}, where buyer might overlook several contextual factors for reformulation caused by focusing on previous queries and items viewed. Following techniques are designed to eliminate such bias by capturing cross-sectional pairs.  

\paragraph{Cross-session co-engaged buyer queries:} 
We address \textit{contextual bias} by incorporating cross-session, query reformulation patterns within the dataset. This bucket introduces some diversity while preserving a connection to the buyer’s original query. Training data is derived from cross-session query pairs, i.e., queries from different user sessions where items engaged under one query (source query) are also co-engaged under another query (target query). 

\paragraph{Cross-session 1-hop co-clicked buyer queries:}
The reformulations with this technique inspires exploration and expand the buyer’s search horizon. Training data is derived from \textit{two-Hop neighbors} Query pairs that are connected via co-clicked items in a two-hop relationship in the session-SRP neighborhood. These pairs allow the model to capture exploratory behaviors among buyers, guiding them towards related broader discovery. Figure~\ref{FIG:CROSS-SESSION} provides a quick illustration into the novel technique of mining cross-session buyer query rewrites.


\subsection{Mapping reformulations to Intent buckets}
The previous subsection provides a holistic approach to capture the buyer query reformulation behavior and historical raw data. This subsection complement the dataset collected with additional post processing steps to filter pairs with predefined intent level. Before proceeding further, we discuss a few domain specific terms. A buyer search query can be typically associated with a product category which is defined as the item category from a pre-specified e-commerce taxonomy. A query is also associated with explicitly specified product aspects or filters like “storage” for an “iPhone” or “size” for a “shoe”. These filters are applied by leveraging the Named Entity Recognition (NER) model to identify the explicit aspects of the buyer query. A few sample queries from the user logs are illustrated in Table~\ref{TAB:DATASET-EXAMPELES}. We define the following intent levels with the definitions provided by domain experts.

\begin{figure}
    \centering
    \includegraphics[width=\linewidth]{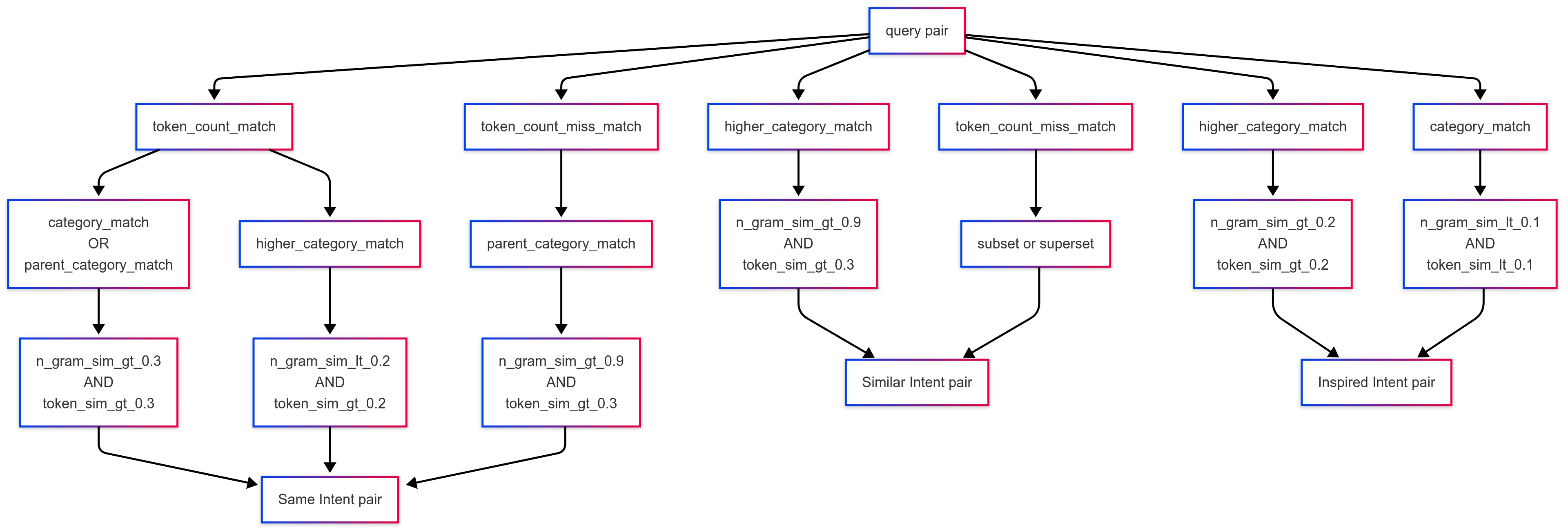}
    \caption{Intent level filtering applied to filter out buyer queries into intent buckets.}    \label{FIG:POST-PROCESS}
\end{figure}

\begin{itemize}
    \item \textbf{Same intent:} This bucket is intended to capture query pairs with high similarity. Here, \textit{nike air jordan 4} and \textit{nike air jordan 11} is a real world example in this bucket. Human judgment by domain experts concluded that token count match does not play a critical role for the following two intent buckets.
    \item \textbf{Similar intent:} The second bucket is intended to capture queries with specificity changes. Here, \textit{nike womens size 9} and \textit{nike womens air max size 9} is a real world example in this bucket. The target query had an additional aspect \textit{air max} specified which make the query more specific. The specificity of source query is broader, whereas the target query is narrower.
    \item \textbf{Inspired intent:} The third bucket captures the queries with low similarity but acts as a pivot which motivates the buyer for relevant products. 
\end{itemize}

\subsection{Application Specific Inference}
Query reformulation as a capability is designed to serve multiple underlying applications as illustrated in Figure~\ref{FIG:OVERVIEW}. Individual applications prefer the reformulation to be specific to the use case. For example, the goal of \textit{NL recovery} module is to enhance relevant recall for the buyer queries with NL results. However, \textit{Advertisement} module focuses on aspect replacement or inspired queries. Assuming individual models for each application tends to suffer from concerns including but not limited to:
\begin{enumerate}
    \item Query reformulation is the underlying problem which these applications solve and this leads to duplication of efforts.
    \item Buyer data driven models requires regular timely updates and multiple pipelines exacerbates this concern further.
    \item Scaling multiple applications becomes cumbersome and causes inefficient use of computational resources.
    \item Implementing global changes as each pipeline needs to be individually updated. This leads to inflexibility in adapting to new technology and practices.
    \item Individual pipelines tend to cause misalignment when the same reformulation is presented in different modules. An efficient system should avoid such redundant queries.
\end{enumerate}

\begin{table*}
    \caption{Frequency of Rewrite types within the test data and model predictions. Highlighted cells represent the best-performing scores in each column.}
    \label{TAB:PERF-BUCKET}
    \begin{tabular}{ccccccccc}
    \toprule
         & \textbf{Empty}   & \textbf{Same}   & \textbf{SuperSet}   & \textbf{SubSet}   & \textbf{Replace}   & \textbf{SubSetRep.}   & \textbf{SupSetRep.}   & \textbf{Other} \\
        
    \midrule
        Test Data          & 0.00    & 0.00   & 53.15      & 7.05     & 39.05     & 0.29  & 0.36   & 0.09    \\
    \midrule
        $\theta_{R}$ & 0.00    & 66.96  & 0.00 & 33.04    & 0.00 & 0.00  & 0.00   & 0.00\\
        $\theta_{S}$  & 0.00    & 38.54  & 0.00       & 14.07 & 25.53 & 6.43  & 8.09   & 7.35    \\
        $\theta_{K}$ & 1.38    & 17.59  & 0.00       & 80.80 & 0.02 & 0.06  & 0.00   & 0.15    \\
    \hdashline
        {$\theta_{B}$}  & 2.37    & 10.90  & 13.80      & 13.75 & 25.41 & 9.59  & 5.92   & 18.26   \\
        {$\theta_{T}$}   & 0.51    & \hspace{-0.1cm}\colorbox{green!30}{1.17}   & \hspace{-0.1cm}\colorbox{green!30}{18.49} & \hspace{-0.1cm}\colorbox{green!30}{7.68} & \hspace{-0.1cm}\colorbox{green!30}{34.23} & 7.29  & 12.73  & 17.91   \\
    \bottomrule
    \end{tabular}
\end{table*}

\begin{table*}
    \caption{Performance evaluation of models trained on datasets generated using different approaches. Highlighted cells represent the best-performing scores in each column.}
    \label{TAB:PERF-EVAL}
    \centering
    \resizebox{\textwidth}{!}{%
    \begin{tabular}{lccccc|ccc}
        \toprule
        & \textbf{cov.} & \textbf{rec. (Sb, Rp, Sp, SbRp, SpRp)} & \textbf{pre. (Sb, Rp, Sp, SbRp, SpRp)} & \textbf{bleu} & \textbf{rougeL} & \textbf{rats} & \textbf{rtfw\_rec.} & \textbf{rtfw\_pre.} \\
        \midrule
        $\theta_{R}$ & 0.33 & 0.15 (0.44, 0, 0, 0, 0) & 0.23 (0.44, 0, 0, 0, 0) & 20.08 & \colorbox{green!30}{0.63} & 0.03 & 0.15 & 0.23 \\
        $\theta_{S}$ & 0.61 & 0.24 (0.42, 0.46, 0, 0.36, \colorbox{green!30}{0.47}) & 0.33 (0.42, 0.46, 0, 0.36, \colorbox{green!30}{0.47}) & 18.29 & 0.55 & 0.13 & 0.24 & 0.33 \\
        $\theta_{K}$ & 0.82 & 0.37 (0.46, 0.33, 0, 0.29, 0) & \colorbox{green!30}{0.70} (0.46, 0.33, 0, 0.29, 0) & 21.82 & 0.58 & 0.07 & 0.37 & \colorbox{green!30}{0.70} \\
        \hdashline
        $\theta_{B}$ & 0.89 & 0.32 (0.47, 0.48, 0.60, 0.31, 0.39) & 0.41 (0.47, 0.48, 0.60, 0.31, 0.39) & 18.55 & 0.41 & 0.40 & 0.32 & 0.41 \\
        $\theta_{T}$ & 0.99 & 0.39 (0.50, 0.49, 0.61, 0.32, 0.39) & 0.44 (\colorbox{green!30}{0.50}, 0.49, 0.61, 0.32, 0.39) & 18.74 & 0.41 & \colorbox{green!30}{0.51} & 0.39 & 0.44 \\
        $\theta_{T}@5$ & \colorbox{green!30}{1.00} & \colorbox{green!30}{0.45} (0.48, \colorbox{green!30}{0.60}, \colorbox{green!30}{0.62}, \colorbox{green!30}{0.42}, 0.42) & 0.54 (0.48, \colorbox{green!30}{0.60}, \colorbox{green!30}{0.62}, \colorbox{green!30}{0.42}, 0.42) & \colorbox{green!30}{22.21} & 0.47 & 0.48 & \colorbox{green!30}{0.45} & 0.53 \\
        \bottomrule
    \end{tabular}
    }
\end{table*}

\section{Performance Evaluation}
\label{SEC:EXPERIMENTS}

We evaluate the effectiveness of the proposed data generation framework by training and comparing multiple models, including both heuristic and learned approaches. The heuristic baseline $\theta_R$ is a stochastic token drop model that randomly deletes up to 50\% of tokens for queries with more than three tokens. This serves as a minimal-effort baseline for evaluating the learning signal in the training data.

The $\theta_S$ model is a BiLSTM trained exclusively on \textit{In-Session} buyer reformulation data, capturing patterns primarily observed in short, corrective user reformulations. $\theta_K$ incorporates structured Knowledge Graph features to guide the model toward token drop behaviors prevalent in entity-aware refinements. In contrast, $\theta_B$ and $\theta_T$ (BiLSTM and lightweight Transformer, respectively) are trained on the dataset generated through our proposed mining pipeline, which unifies reformulation behaviors across sessions and product spaces. The $\theta_T@5$ variant evaluates the top-5 rewrites generated by the Transformer model, offering a proxy for beam-search style outputs in production use cases.

Our evaluation is threefold. \textit{First}, we perform a rewrite-type frequency alignment analysis to compare the structural reformulation patterns captured by the models. \textit{Second}, we use standard machine translation-style metrics to assess token-level fidelity. \textit{Third}, we propose domain-specific evaluation metrics designed to expose failure cases and misalignments that traditional metrics may overlook. Metrics with highest performance values in each column are highlighted in \textit{green} to facilitate comparison.

\subsection{Rewrite Type Analysis}
Table~\ref{TAB:PERF-BUCKET} presents a comparative analysis of the distribution of rewrite types generated by each model relative to the ground-truth annotations. The columns capture key reformulation categories: \textit{Empty} for no output, \textit{Same} for identical queries, \textit{SuperSet} for token additions, \textit{SubSet} for deletions, \textit{Replace} for token substitutions, and the combinations \textit{SubSetRep.} and \textit{SupSetRep.} for compound behaviors.

The highlighted entries denote model distributions most closely aligned with those observed in the test set. Notably, $\theta_T$ demonstrates broad coverage across all rewrite types, suggesting its generalization across diverse user intents and downstream task expectations. In contrast, the deviation of $\theta_S$ and $\theta_K$ from the expected distributions underscores their bias toward their respective data sources—short in-session corrections and entity-guided minimal rewrites, respectively. These deviations are particularly relevant in vertical-specific use cases such as Null and Low-frequency (N\&L) query recovery, where targeted behaviors like entity completion or disambiguation dominate.

\subsection{Standard Metrics}
Table~\ref{TAB:PERF-EVAL} reports standard metrics including coverage (\textit{cov.}), token-level recall and precision (with granular breakdowns by rewrite type), and widely-used translation scores such as \textit{BLEU} and \textit{ROUGE-L}.

Here, \textit{rec.} denotes the proportion of target tokens captured by the model, while \textit{pre.} measures the correctness of predicted tokens relative to the gold rewrite. The breakdown across Subset (Sb), Replace (Rp), Superset (Sp), and their combinations (SbRp, SpRp) enables a nuanced view of how well each model captures specific transformation types. $\theta_K$ demonstrates high precision, largely attributable to its focus on deletion-based behavior. $\theta_T$ and $\theta_T@5$ strike a more balanced tradeoff between recall and precision, especially in substitution-heavy or compound rewrites.

While BLEU and ROUGE-L provide comparative benchmarks, their sensitivity to word order and n-gram overlap limits their utility in high-variance reformulations such as product search rewrites. This motivates the introduction of more targeted metrics.

\subsection{Rewrite Type–Focused Metrics}
To address the limitations of conventional metrics, we introduce the \textit{Rewrite Type Agreement Score} (RATS), a novel metric designed to assess the structural fidelity of rewrites. RATS measures the proportion of model predictions that match the rewrite type of the ground-truth reformulation:

\[
\text{rats} = \frac{1}{N} \sum_{i=1}^{N} \mathbf{1}[\text{rewrite\_type}(\hat{y}_i) = \text{rewrite\_type}(y_i)]
\]

where $N$ is the number of instances, $y_i$ is the reference reformulation, and $\hat{y}_i$ is the model prediction. Additionally, we introduce \textit{Rewrite Type Frequency–Weighted Recall and Precision} (\textit{rtfw\_rec.}, \textit{rtfw\_pre.}), which weight the recall and precision scores by the frequency distribution of rewrite types in the test set. These metrics provide a holistic evaluation that favors models performing well on both common and edge-case rewrite patterns. $\theta_T$ and $\theta_T@5$ exhibit consistently strong performance across RATS and type-weighted metrics, reflecting their robustness across user-intent distributions. These results validate the effectiveness of the mined training data in capturing a representative sample of real-world reformulation behavior.

\section{Conclusion and Future Work}

This study explores buyer query reformulation behavior by mining in-session and cross-session queries, using co-clicked items to classify intents. We use this dataset to train a Neural Machine Translation (NMT) model capable of providing multiple intent-specific query reformulations. The model’s holistic evaluation reveals its potential to improve the buyer journey in e-commerce by enhancing query relevance and user engagement.

Future work includes refining the model to handle ambiguous queries and rare reformulation types, potentially by incorporating additional contextual data such as browsing history or external product metadata. Integrating Retrieval-Augmented Generation (RAG) systems, which combine retrieval-based and generative approaches, could further enhance query reformulations by leveraging external knowledge. Finally, expanding to multimodal inputs like images or voice search could broaden the model's applicability to diverse e-commerce environments, offering more intuitive and flexible search experiences.

In summary, this work advances e-commerce search by improving query reformulation and paves the way for future innovations in intelligent search and recommendation systems.


\end{document}